\documentclass[fleqn,twoside]{article}
\makeatletter
\def\@fnsymbol#1{\ensuremath{\ifcase#1\or *\or **\or 
   \mathsection\or \mathparagraph\or \|\or \else\@ctrerr\fi}}
\makeatother
\usepackage{espcrc2}

\usepackage{graphicx}
\usepackage[figuresright]{rotating}

\newcommand{\AmS}{{\protect\the\textfont2
  A\kern-.1667em\lower.5ex\hbox{M}\kern-.125emS}}

\title{
Preliminary study of metabolic radiotherapy with $^{188}$Re via small animal imaging
}

\author{A. Antoccia\address[biorm3]{Dept. of Biology, Univ. Roma3, V.le G. Marconi, I-00146 Roma}\address[infnrm3]{INFN, Sezione Roma3, Via della Vasca Navale 84, I-00146 Roma}, 
        G. Baldazzi\address[fisbo]{Dept. of Physics, Univ. Bologna, V.le C. Berti-Pichat 6/2, I-40127 Bologna}\address[infnbo]{INFN, Sezione Bologna, V.le C. Berti-Pichat 6/2, I-40127 Bologna},
        M. Bello\address[fispd]{Dept. of Physics, Univ. Padova, Via F. Marzolo 8, I-35131 Padova}\address[lnl]{INFN - LNL, V.le dell'Universit\`a 2, I-35020 Legnaro},
        D. Bernardini\address[vetpd]{Dip. Scienze Cliniche Veterinarie, Univ. Padova, c/o Agripolis, V.le dell'Universit\`a 16, I-35020 Legnaro}
        P. Boccaccio\addressmark[lnl],
        D. Bollini\addressmark[fisbo]\addressmark[infnbo], 
        F.~de~Notaristefani\address[fisrm3]{Dept. of Physics, Univ. Roma3, Via della Vasca Navale 84, I-00146 Roma}\addressmark[infnrm3],
        F. Garibaldi\address[iss]{Ist. Superiore di Sanit\`a, V.le Regina Elena 299, I-00161 Roma} 
        G. Hull\addressmark[fisrm3]\addressmark[infnrm3],
        U.~Mazzi\address[scfarmpd]{Dept. of Pharm. Sc., Univ. Padova, Via F. Marzolo 5, I-35131 Padova},
        G. Moschini\addressmark[fispd]\addressmark[lnl],
        A.~Muciaccio\address[facfarmbo]{Faculty of Pharmacy, Univ. Bologna, Via S. Donato 19/2, I-40126 Bologna}\thanks{present address: Via Parigi 5, I-06019 Umbertide}, 
        F.-L.~Navarria\addressmark[fisbo]\addressmark[infnbo]\thanks{e-mail: navarria@bo.infn.it},
        V. Orsolini Cencelli\addressmark[infnrm3],
        G.~Pancaldi\addressmark[infnbo],
        R. Pani\address[fisrm1]{Dip. Medicina Sperimentale e Patologia, Univ. Roma1, V.le Regina Elena 324, I-00161 Roma}\address[infnrm1]{INFN, Sezione Roma1, P.le Aldo Moro 2, I-00185 Roma}
        A.~Perrotta\addressmark[infnbo],
        M.~Riondato\addressmark[scfarmpd], 
        A.~Rosato\address[sconcpd]{Dept. of Oncological and Surgical Sc., Univ. of Padova, Via Gattamelata 64, I-35128 Padova}, 
        A.~Sgura\addressmark[biorm3]\addressmark[infnrm3],
        C.~Tanzarella\addressmark[biorm3]\addressmark[infnrm3],
        N. Uzunov\addressmark[lnl]\address[shumen]{Dept. of Natural Sciences, ``K. Preslavsky'' Univ. of Shumen, Bulgaria}
        and         
        M. Zuffa\addressmark[infnbo]
        } 
       
\begin{document}

\begin{abstract}
$^{188}$Re is a $\beta^-$ (Emax = 2.12 MeV) and $\gamma$ (155 keV) emitter. 
Since its 
chemistry is similar to that of the largely employed tracer, $^{99m}$Tc, 
 molecules of hyaluronic acid  (HA) have been labelled with $^{188}$Re
to produce a target specific radiopharmaceutical.
The radiolabeled compound, i.v. injected in healthy mice, is able to 
accumulate into the liver after a few minutes.
To study the effect of metabolic radiotherapy in mice, 
we have built a small gamma camera based on a matrix of YAP:Ce crystals, 
with 0.6$\times$0.6$\times$10 mm$^3$ pixels, read out by a R2486 Hamamatsu PSPMT. A 
high-sensitivity 20 mm 
thick lead parallel-hole collimator, with hole diameter 1.5 mm and septa 
of 0.18 mm, is placed in front of the YAP matrix. 
Preliminary results obtained with 
various phantoms containing a solution of $^{188}$Re and with C57 black mice 
injected with the $^{188}$Re-HA solution are presented. 
To increase the space resolution and to obtain two orthogonal projections 
simultaneously we are building in parallel two new 
cameras to be positioned at 90 degrees. They use a CsI(Tl) matrix with 
1$\times$1$\times$5 mm$^3$ pixels read out by H8500 Hamamatsu Flat panel PMT. 
\vspace{1pc}
\end{abstract}

\maketitle

\section{INTRODUCTION}

$^{188}$Re is an attractive therapeutic radioisotope with broad clinical 
applications. Oncology applications range from palliation of metastatic 
bone pain to bone marrow ablation 
and in general to the use of $^{188}$Re labelled therapeutic agents to 
target specific cancerous tissues (see for instance reference \cite{iaea}).
Other therapeutic applications comprise the inihibition of restenosis 
after Percutanueous Transluminal Coronary Angioplasty (PTCA), 
radiation synovectomy,
and intravasal brachytherapy. 
In the field of metabolic radiotherapy  
$^{188}$Re shows several 
favourable characteristics. It can be produced carrier free using a 
W-Re generator (see for instance reference \cite{ornl}) and its chemistry 
is similar to that of $^{99m}$Tc which is the most used radioisotope in 
nuclear medicine (medical imaging). 
The $^{188}$W parent has a 69 d half-life, which permits to use 
the generator for a relatively long period, the equilibrium between 
parent and daughter setting up within about two days. 
$^{188}$Re decays to $^{188}$Os$^*$ in about 0.7 
days via the emission of a $\beta$-ray with a maximum energy of 2.12 MeV 
(0.78 MeV average energy), 
which can be used for destroying cancerous cells. In addition $^{188}$Os$^*$
emits promptly (0.69 ns) a 155 keV $\gamma$-ray (15\%), which can be used for 
imaging. Given the chemical similarity with $^{99m}$Tc, it can be linked 
to molecules of hyaluronic acid (HA) which have the function of carrying 
it to specific sites in the body, e.g. with the production of an 
accumulation and retention of the drug in the liver \cite{sirr}. 
Hence the potential interest for treating liver cancers. 
On the other hand the $\beta$ radiation could interfere with the labelling 
process and destroy the molecule used to carry the radioisotope in the body 
to the target organ, thus reducing the therapeutic effect. Two additional 
points have to do with the 
relatively long lifetime and with the rich photon spectrum of $^{188}$Re, 
which extends to high energy, compared to the single 140 keV photon emitted 
by $^{99m}$Tc with a half-life of about 6 h. Taking into account the 
branching ratios (BR) and the lifetimes, the relative $^{188}$Re/$^{99m}$Tc 
counting rate, for an equal number of $\mu$-moles, is about 9\%. This 
implies that photon detectors developed for imaging with $^{99m}$Tc are 
not necessarily sensitive enough. The usual thickness of a 
low-energy Pb collimator, 
say 20 mm, will be almost transparent to $^{188}$Re photon lines at 300 keV or 
higher energies, even with BRs depressed by a factor of ten or more, 
producing background counts which blur the image and worsen the spatial 
resolution. To study the effect of metabolic radiotherapy in mice, we have 
therefore built a new small high-sensitivity $\gamma$-camera, following the 
experience with the YAP-camera \cite{yap}\cite{yap1}\cite{yap2}, which is used routinely to 
image small animals (mice) 
with  $^{99m}$Tc-HA at the Laboratori Nazionali di Legnaro, Italy (e.g. \cite{habut}\cite{99mtc}). 
  
\section{THE GAMMA CAMERA}

The gamma camera is based on a matrix of yttrium aluminium perovskite doped 
with cerium (YAP:Ce or YAlO$_3$:Ce) crystals \cite{preciosa}, with 
friendly mechanical properties (no igroscopicity), fast response ($\sim$25 ns 
decay time),   
high density (5.37 g/cm$^3$) and good X- and $\gamma$-ray absorption. 
There are 66$\times$66 pixels, each 0.6$\times$0.6$\times$10 mm$^3$, covering a field-of-view 
(FOV) of 
40$\times$40 mm$^2$. The pixels are covered laterally by a special 5 $\mu$m thick 
reflective coating which provides also the optical separation between 
neighboring elements. The scintillator is read out by a R2486 Hamamatsu 
PositionSensitivePMT \cite{hamamatsu} with a 76 mm diameter photocathode.
The anode consists of 16 plus 16 wires crossing at 90$^o$ and connected by 
two resistive chains, defining the $x$ and $y$ directions. The wires define 
an active area with a diameter of about 50 mm.   
A 20 mm thick lead parallel hexagonal-hole collimator \cite{nuclfields}, 
with hole diameter 1.5 mm and septa of 0.18 mm, is placed in front of the 
YAP matrix. The detector is triggered 
using the last dynode and the ends of the $x$ and $y$ resistive chains 
(x$_1$, x$_2$, y$_1$, y$_2$) are amplified, stretched and read out 
by a PC using 
a PCI 6023E card \cite{national}. The coordinates of the photon impact 
point are then reconstructed by charge division, 

\begin{equation}
x = {\rm {(x_1 - x_2)}}/{\rm {(x_1 + x_2)}}
\end{equation}
 
\noindent and similarly for $y$. 

\section{CALIBRATIONS}

\begin{figure}[htb]
\vspace{9pt}
\includegraphics[width=6.9 truecm]{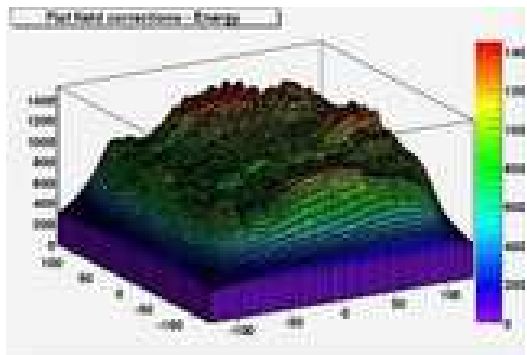}
\vspace{-1.5pc}
\caption{Flat field calibration with $^{99m}$Tc.} 
\label{fig:flat}
\vspace{-1pc}
\end{figure}

The energy response of the detector to a spatially uniform source (flat 
field) of $^{99m}$Tc 140 keV photons prior to energy equalization has 
been determined using a solution containing $^{99m}$Tc which covered the 
whole FOV and was located a few centimeters in front of the collimator 
surface. The $xy$ has been arbitrarily divided in pixels and the 
average energy computed in each pixel.  The corrections to the measured 
energy extracted from this 
calibration are shown in Fig.~\ref{fig:flat} and have been used in the 
following. As shown below with the $^{188}$Re spectrum, these 
corrections improve the energy resolution of the detector. Apart from 
the energy equalization, no other correction has been applied. 
With $^{99m}$Tc the sensitivity of the gamma camera is found to be 
$\sim$2$\times$10$^{-4}$ or $\sim$7$\times$10$^3$ cps/mCi, which 
agrees roughly with calculations.

\begin{figure}[htb]
\vspace{9pt}
\includegraphics[width=6.9 truecm]{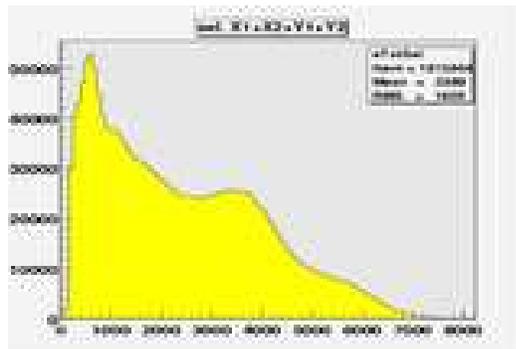}
\vspace{-1.5pc}
\caption{$^{241}$Am, $^{57}$Co and $^{137}$Cs superimposed spectra measured with the YAP camera.}
\label{fig:acc-en}
\vspace{-1pc}
\end{figure}

\begin{figure}[htb]
\vspace{9pt}
\includegraphics[width=6.9 truecm]{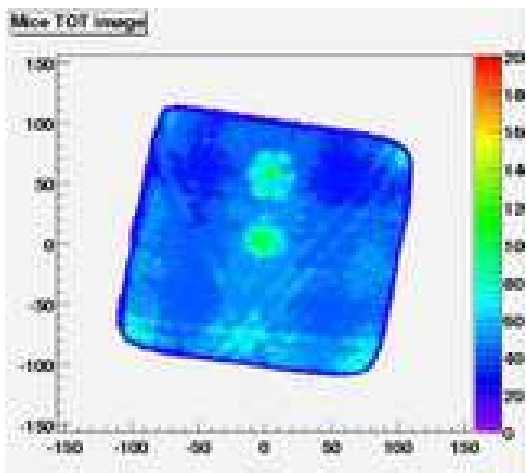}
\vspace{-1.5pc}
\caption{$^{241}$Am (center), $^{57}$Co (top) and $^{137}$Cs (bottom) images measured with the YAP camera.}
\label{fig:acc-im}
\vspace{-1pc}
\end{figure}

Pointlike $^{241}$Am (60 keV photons), $^{57}$Co ($\sim$122 keV photons) 
and $^{137}$Cs (660 keV) sources located 
in different positions a few millimeters distant from the collimator 
have been used  
both to simulate the $^{188}$Re energy spectrum and to evaluate 
the spatial resolution of the detector. In addition $^{241}$Am and $^{57}$Co, 
together with $^{99m}$Tc photons, permit the calibration of energy 
scale. 
The overall energy spectrum is 
presented in Fig.~\ref{fig:acc-en} and the cumulative image of the sources is 
visible in 
Fig.~\ref{fig:acc-im}. The 60 keV is well prominent and a 122 keV shoulder 
is also visible; $^{137}$Cs instead produces a broad shoulder at about 
half of the photon energy.  
Using appropriate energy cuts the three images 
can be separated. 
The intrinsic spatial resolution of the system is quite good, since 
individual collimator holes can be clearly seen in Fig.~\ref{fig:acc-im}. The 
resolution however is worsened in practice by the small thickness of the septa 
relative to the hole diameter of the present collimators, making them 
partially transparent to radiation, so the actual effective  
resolution is $\Delta$$x$ $\sim$ 3 mm FWHM at 122 keV.

\begin{figure}[htb]
\vspace{9pt}
\includegraphics[width=8.2 truecm]{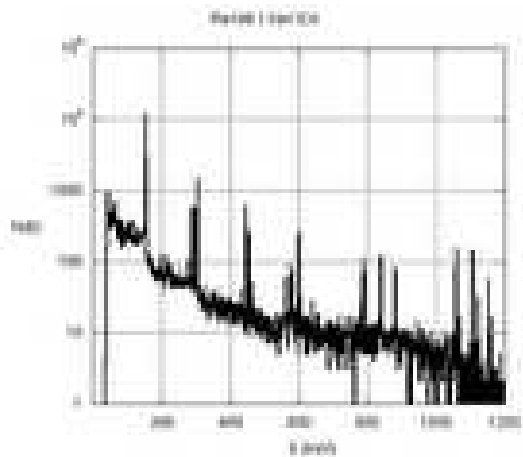}
\vspace{-2pc}
\caption{$^{188}$Re spectrum measured with a Ge detector.}
\label{fig:re-ge}
\vspace{-1pc}
\end{figure}

The $^{188}$Re photon spectrum measured with a Ge detector is shown in
Fig.~\ref{fig:re-ge}. The 155 keV line is prominent, but many more lines 
are present at higher energy, in some cases, e.g. at $\sim$300 keV, with 
BRs only a factor of ten lower, or at 800-1200 keV with intensities 
lower only by a factor of 100 \cite{nndc}. 

\begin{figure}[htb]
\vspace{9pt}
\includegraphics[width=6.9 truecm]{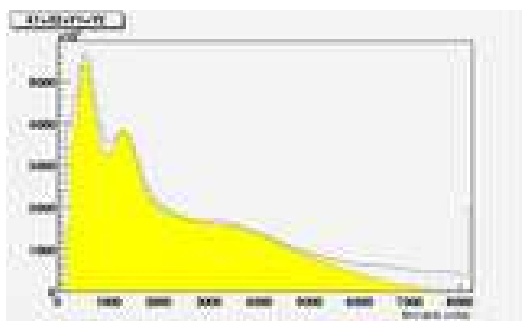}
\vspace{-1.5pc}
\caption{$^{188}$Re spectrum measured with the YAP camera: the shaded spectrum has been corrected for energy non-uniformity.}
\label{fig:re-yap}
\vspace{-1pc}
\end{figure}

The $^{188}$Re spectrum measured with the YAP camera 
during the imaging of a C57 mouse (see next section) 
is shown in Fig.~\ref{fig:re-yap}. The spectrum is shown before and after 
applying the corrections for energy non-uniformity, and a clear improvement 
is observed with the shrinking of the 155 keV peak. After the corrections,
the energy resolution is $\Delta$E/E $\sim$ 33\% $@$ 155 keV. 

\section{FIRST MEASUREMENTS WITH $^{188}$Re}

The labelling reaction of HA using $^{188}$Re
was carried out with good 
yields (65-70\%). 
The radiolabelled compound was purified with a size exclusion
chromatographic method before being used for biodistribution studies. 
Stability studies in rat serum confirmed the maintaining  of the Re linked 
to the polymer and there was no evidence of radio-decomposition after a few 
hours. 

The radiotoxicity of $^{188}$Re has been tested ``in vitro''  and compared 
with $^{99m}$Tc, with which no effect is expected.  Cells of the 
M5076 tumor line have been treated with $^{188}$Re and  $^{99m}$Tc solutions,
and irradiated with X-rays. The number of binucleate cells and of micronuclei 
in the cells is then counted. 
M5076 cells turn to out to be highly sensitive to X-rays (0.25-2 
Gy)\cite{sirr}. Activities of 150-300 $\mu$Ci of $^{99m}$Tc show no effect, 
but $^{188}$Re $\beta$-rays, with similar initial activities integrated 
during 72 h, seem to be quite efficient in inducing DNA damage\cite{sif}. 

\begin{figure}[htb]
\vspace{9pt}
\includegraphics[width=6.9 truecm]{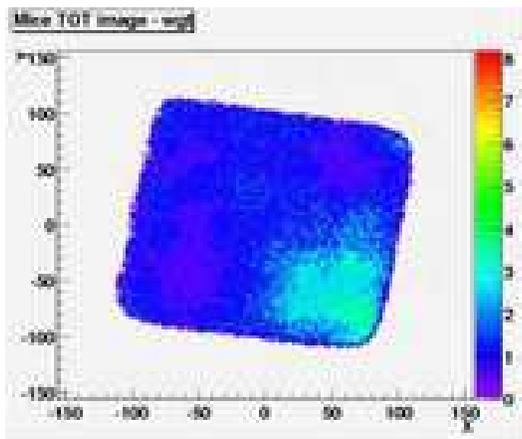}
\vspace{-1.5pc}
\caption{The image of the C57 mouse integrated for the first 5 minutes after the injection of $^{188}$Re-HA in the caudal vein.}
\label{fig:c57a}
\vspace{-1pc}
\end{figure}

\begin{figure}[htb]
\vspace{9pt}
\includegraphics[width=6.9 truecm]{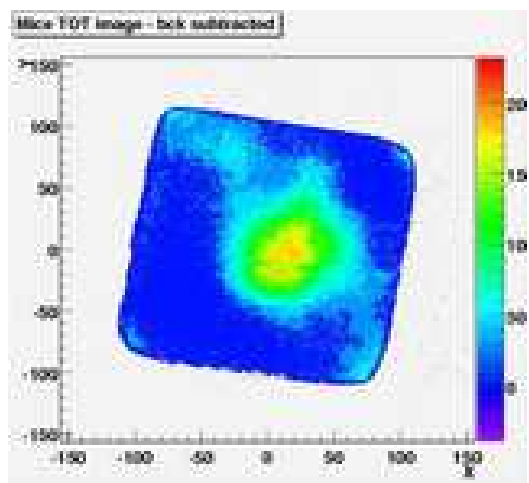}
\vspace{-1.5pc}
\caption{The image of the C57 mouse integrated between 5 and 185 minutes after the injection of $^{188}$Re-HA in the caudal vein. The volume of large activity corresponds to the liver.}
\label{fig:c57b}
\vspace{-1pc}
\end{figure}

\begin{figure}[htb]
\vspace{9pt}
\includegraphics[width=6.9 truecm]{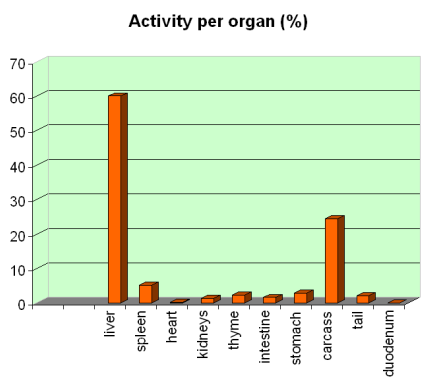}
\vspace{-1.5pc}
\caption{Activity of various organs ``$post$ $mortem$''.}
\label{fig:att}
\vspace{-1pc}
\end{figure}

To test the full chain, from the radiolabelling to to the imaging ``in vivo'',
a C57 black mouse (healthy, female) has been injected with  $^{188}$Re-HA.
After general anesthesia, the solution with an activity of about 
250 $\mu$Ci was injected in the caudal vein.
The mouse was positioned along the diagonal of the FOV, with the locus of 
injection outside it, and was monitored for about three hours.
The  image collected in the first five minutes shows a large spot close to 
the locus of injection in the tail (Fig.~\ref{fig:c57a}). After 5 mins the 
activity concentrates roughly in the centre of the body, in a volume which 
contains the liver (Fig.~\ref{fig:c57b}). The activity is slowly 
decreasing during the 3 h of the measurement. After 3 h the mouse was 
sacrificed, and the organs were extracted and measured with a microcurimeter
(Fig.~\ref{fig:att}). The liver contains 60\% of the residual activity 
and close by organs another 20\%, in agreement with the scintigrafic 
image (Fig.~\ref{fig:c57b}), where individual organs are not 
resolved. Even with limited resolution the test shows that it is 
possible to monitor the biodistribution of $^{188}$Re in mice, with a 
potential saving in the number of animals needed for testing the 
$^{188}$Re therapy. 

\section{CONCLUSIONS}

Preliminary results obtained using a new YAP camera in imaging 
$^{188}$Re sources
and C57 black mice 
injected with a $^{188}$Re-HA solution have been presented. 
To increase the space resolution without 
losing sensitivity, and to obtain different projections simultaneously, 
we are building two new 
cameras to be positioned at 90 degrees around the animal. 
They use a CsI(Tl) matrix with 1$\times$1$\times$5 
mm$^3$ pixels read out by H8500 Hamamatsu Flat panel PMT \cite{hamamatsu}. 
Parallel-hole Pb 
collimators 20 mm thick, with 1 mm diameter hexagonal holes and 0.2 mm 
thick septa, will be mounted in front of the scintillators. 
Also specially made collimators with thicker septa and/or different 
absorber material will be used. The front-end 
electronics for the 64 channels of the H8500 has been designed using 
MPX-08 chips \cite{nova}. The system will be mounted on a rotating support 
in order to produce tomographic images.

\end{document}